\documentclass[aps,amsmath,amssymb,twocolumn,superscriptaddress,nofootinbib,showpacs]{revtex4-1}

\usepackage{graphicx}
\usepackage{dcolumn}
\usepackage{bm}
\usepackage{color}
\usepackage[normalem]{ulem}
\usepackage[colorlinks=true,pdfstartview=FitV,linkcolor=blue,citecolor=blue,urlcolor=blue]{hyperref}
\usepackage[mathlines]{lineno}

\everymath{\displaystyle}
\begin{document}

\newcommand{\up}[1]{$^{#1}$}
\newcommand{\down}[1]{$_{#1}$}
\newcommand{\powero}[1]{\mbox{10$^{#1}$}}
\newcommand{\powert}[2]{\mbox{#2$\times$10$^{#1}$}}

\newcommand{\mchi}{\mbox{$m_\chi$}}
\newcommand{\dedx}{$\rm{dE/dx}$}
\newcommand{\gev}{\mbox{GeV\,}$c^{-2}$}
\newcommand{\swn}{\mbox{$\sigma_{\chi-n}$}}
\newcommand{\evr}{\mbox{eV$_{\rm nr}$}}
\newcommand{\eve}{\mbox{eV$_{\rm ee}$}}
\newcommand{\dru}{\mbox{keV$_{\rm ee}^{-1}$\,kg$^{-1}$\,d$^{-1}$}}
\newcommand{\um}{\mbox{$\mu$m}}
\newcommand{\sxy}{\mbox{$\sigma_{xy}$}}
\newcommand{\sx}{\mbox{$\sigma_{x}$}}
\newcommand{\smax}{\mbox{$\sigma_{\rm max}$}}
\newcommand{\spix}{\mbox{$\sigma_{\rm pix}$}}
\newcommand{\obo}{\mbox{1$\times$1}}
\newcommand{\obh}{\mbox{1$\times$100}}
\newcommand{\dll}{\mbox{$\Delta LL$}}
\newcommand*\diff{\mathop{}\!\mathrm{d}}
\newcommand*\Diff[1]{\mathop{}\!\mathrm{d^#1}}

\newcommand{\tritium}{\mbox{$^{3}$H}}
\newcommand{\ironfive}{\mbox{$^{55}$Fe}}
\newcommand{\coseven}{\mbox{$^{57}$Co}}
\newcommand{\pbten}{$^{210}$Pb}
\newcommand{\biten}{$^{210}$Bi}
\newcommand{\sitwo}{$^{32}$Si}
\newcommand{\ptwo}{$^{32}$P}
\newcommand{\Ez}{\mbox{$E\text{-}z$} }
\newcommand{\Esigma}{\mbox{$E\text{-}\sigma_x$} }

\newcommand{\comment}[1]{\textcolor{blue}{#1}}

\title{Results on Low-Mass Weakly Interacting Massive Particles from an 11\,kg\,d Target Exposure of DAMIC at SNOLAB}

\author{A.~Aguilar-Arevalo} 
\affiliation{Universidad Nacional Aut{\'o}noma de M{\'e}xico, Mexico City, Mexico} 

\author{D.~Amidei}
\affiliation{Department of Physics, University of Michigan, Ann Arbor, Michigan, USA}  

\author{D.~Baxter}
\affiliation{Kavli Institute for Cosmological Physics and The Enrico Fermi Institute, The University of Chicago, Chicago, Illinois, USA}

\author{G.~Cancelo}
\affiliation{Fermi National Accelerator Laboratory, Batavia, Illinois, USA}

\author{B.A.~Cervantes Vergara}
\affiliation{Universidad Nacional Aut{\'o}noma de M{\'e}xico, Mexico City, Mexico} 

\author{A.E.~Chavarria}
\affiliation{Center for Experimental Nuclear Physics and Astrophysics, University of Washington, Seattle, Washington, USA}

\author{ J.C.~D'Olivo}
\affiliation{Universidad Nacional Aut{\'o}noma de M{\'e}xico, Mexico City, Mexico} 

\author{J.~Estrada}
\affiliation{Fermi National Accelerator Laboratory, Batavia, Illinois, USA}

\author{F.~Favela-Perez}
\affiliation{Universidad Nacional Aut{\'o}noma de M{\'e}xico, Mexico City, Mexico} 

\author{R.~Ga\"ior}
\affiliation{Laboratoire de Physique Nucl\'eaire et des Hautes \'Energies (LPNHE), Sorbonne Universit\'e, Universit\'e de Paris, CNRS-IN2P3, Paris, France}

\author{Y.~Guardincerri}
\thanks{Deceased January 2017}
\affiliation{Fermi National Accelerator Laboratory, Batavia, Illinois, USA}

\author{E.W.~Hoppe}
\affiliation{Pacific Northwest National Laboratory (PNNL), Richland, Washington, USA} 

\author{ T.W.~Hossbach}
\affiliation{Pacific Northwest National Laboratory (PNNL), Richland, Washington, USA} 

\author{B.~Kilminster}
\affiliation{Universit{\"a}t Z{\"u}rich Physik Institut, Zurich, Switzerland }

\author{I.~Lawson}
\affiliation{SNOLAB, Lively, Ontario, Canada }

\author{S.J.~Lee}
\affiliation{Universit{\"a}t Z{\"u}rich Physik Institut, Zurich, Switzerland }

\author{A.~Letessier-Selvon}
\affiliation{Laboratoire de Physique Nucl\'eaire et des Hautes \'Energies (LPNHE), Sorbonne Universit\'e, Universit\'e de Paris, CNRS-IN2P3, Paris, France}

\author{A.~Matalon}
\affiliation{Kavli Institute for Cosmological Physics and The Enrico Fermi Institute, The University of Chicago, Chicago, Illinois, USA}
\affiliation{Laboratoire de Physique Nucl\'eaire et des Hautes \'Energies (LPNHE), Sorbonne Universit\'e, Universit\'e de Paris, CNRS-IN2P3, Paris, France}

\author{P.~Mitra}
\affiliation{Center for Experimental Nuclear Physics and Astrophysics, University of Washington, Seattle, Washington, USA}

\author{C.T.~Overman}
\affiliation{Pacific Northwest National Laboratory (PNNL), Richland, Washington, USA} 

\author{A.~Piers}
\affiliation{Center for Experimental Nuclear Physics and Astrophysics, University of Washington, Seattle, Washington, USA}

\author{P.~Privitera}
\affiliation{Kavli Institute for Cosmological Physics and The Enrico Fermi Institute, The University of Chicago, Chicago, Illinois, USA}
\affiliation{Laboratoire de Physique Nucl\'eaire et des Hautes \'Energies (LPNHE), Sorbonne Universit\'e, Universit\'e de Paris, CNRS-IN2P3, Paris, France}

\author{K.~Ramanathan}
\affiliation{Kavli Institute for Cosmological Physics and The Enrico Fermi Institute, The University of Chicago, Chicago, Illinois, USA}

\author{J.~Da~Rocha}
\affiliation{Laboratoire de Physique Nucl\'eaire et des Hautes \'Energies (LPNHE), Sorbonne Universit\'e, Universit\'e de Paris, CNRS-IN2P3, Paris, France}

\author{Y.~Sarkis}
\affiliation{Universidad Nacional Aut{\'o}noma de M{\'e}xico, Mexico City, Mexico}

\author{M.~Settimo}
\affiliation{SUBATECH, CNRS-IN2P3, IMT Atlantique, Universit\'e de Nantes, Nantes, France}

\author{R.~Smida}
\affiliation{Kavli Institute for Cosmological Physics and The Enrico Fermi Institute, The University of Chicago, Chicago, Illinois, USA}

\author{R.~Thomas}
\affiliation{Kavli Institute for Cosmological Physics and The Enrico Fermi Institute, The University of Chicago, Chicago, Illinois, USA}

\author{J.~Tiffenberg}
\affiliation{Fermi National Accelerator Laboratory, Batavia, Illinois, USA}

\author{M.~Traina}
\affiliation{Laboratoire de Physique Nucl\'eaire et des Hautes \'Energies (LPNHE), Sorbonne Universit\'e, Universit\'e de Paris, CNRS-IN2P3, Paris, France}

\author{R.~Vilar}
\affiliation{Instituto de F\'isica de Cantabria (IFCA), CSIC--Universidad de Cantabria, Santander, Spain}

\author{A.L.~Virto}
\affiliation{Instituto de F\'isica de Cantabria (IFCA), CSIC--Universidad de Cantabria, Santander, Spain}

\collaboration{DAMIC Collaboration}
\noaffiliation

\date{\today}
\preprint{FERMILAB-PUB-20-284-E}

\begin{abstract}
We present constraints on the existence of weakly interacting massive particles (WIMPs) from an 11\,kg\,d target exposure of the DAMIC experiment at the SNOLAB underground laboratory.
The observed energy spectrum and spatial distribution of ionization events with electron-equivalent energies $>$200\,\eve\ in the DAMIC CCDs are consistent with backgrounds from natural radioactivity.
An excess of ionization events is observed above the analysis threshold of 50\,\eve . 
While the origin of this low-energy excess requires further investigation, our data exclude spin-independent WIMP-nucleon scattering cross sections \swn\ as low as \powert{-41}{3}\,cm$^2$ for WIMPs with masses \mchi\ from 7 to 10\,\gev .
These results are the strongest constraints from a silicon target on the existence of WIMPs with \mchi $<$9\,\gev\ and are directly relevant to any dark matter interpretation of the excess of nuclear-recoil events observed by the CDMS silicon experiment in 2013.
\end{abstract}


\maketitle

The DAMIC experiment at SNOLAB employs the bulk silicon of scientific charge-coupled devices (CCDs) to search for ionization signals produced by interactions of particle dark matter from the Milky Way halo.
Weakly interacting massive particles (WIMPs) are leading candidates to constitute the cold dark matter in the Universe~\cite{Kolb:1990vq, *Griest:2000kj, *Zurek:2013wia}.
WIMPs would have characteristic speeds of hundreds of km\,s$^{-1}$ and would scatter elastically with nuclei to produce nuclear recoils~\cite{PhysRevD.31.3059, LEWIN199687}, which generate ionization signals in detector targets.
By virtue of the low noise of the CCDs and the relatively low mass of the silicon nucleus, DAMIC is particularly sensitive to WIMPs with masses \mchi\ in the range 1--10\,\gev .

In 2013, the CDMS Collaboration reported an excess of nuclear-recoil events observed above their background model in their silicon detectors~\cite{Agnese:2013rvf}, which could be attributed to the scattering of WIMPs with \mchi$\sim$9\,\gev .
Although null results from multiple experimental searches are in tension with this interpretation~\cite{xenon1t, *supercdms, darkside50, *cdmslite, *pico60, *cresstiii, *xenon1ts2}, detailed analyses demonstrate the large sensitivity to theoretical assumptions in the comparison of WIMP search results between different nuclear targets~\cite{Witte_2017}.
In this Letter, we explore with the same nuclear target the parameter space that corresponds to the CDMS event excess.

Throughout 2017--2018, DAMIC acquired data for its dark matter search with a tower of seven 16-megapixel CCDs (6.0\,g each) in the SNOLAB underground laboratory.
Each CCD is held in a copper module that slides into slots of a copper box that is cooled to $\sim$140\,K inside a cryostat.
The top module (CCD~1) was made from high-radiopurity copper electroformed by Pacific Northwest National Laboratory~\cite{HOPPE2014116}, and is shielded above and below by two 2.5-cm-thick lead bricks.
The other modules (CCDs~2--7), made from commercial copper, populate the bottom segment of the box.
The box is shielded on all sides by $\sim$20\,cm of lead and 42\,cm of polyethylene to stop environmental $\gamma$ rays and neutrons, respectively.
The innermost 5\,cm of the lead shield and the bricks inside the box are ancient (smelted more than 300\,years ago) and have reduced radiation from \pbten\ ($\tau_{1/2}$$=$22\,y) contamination.
Boiloff from a liquid nitrogen dewar is used to purge the volume around the cryostat from radon, whose level is continuously monitored.
The overburden of the laboratory site (6010\,m water equivalent) suppresses cosmic-muon backgrounds to a negligible level.
Details of the DAMIC infrastructure can be found in Ref.~\cite{Aguilar-Arevalo:2015lvd}.

The DAMIC CCDs were developed by Lawrence Berkeley National Laboratory MicroSystems Lab~\cite{1185186}.
The CCDs are 674$\pm$3\,\um\ thick with an active thickness of 665$\pm$5\,\um\ that is fully depleted by a bias of 70\,V applied to a back-side planar contact.
Ionizing radiation produces free charges (electron-hole pairs) in the active bulk.
The holes are drifted along the direction of the electric field ($-\hat{z}$) and collected on an array of 4116$\times$4128 pixels of size 15$\times$15\,$\mu$m$^{2}$ ($z$$=$0 plane).
Because the drifting charge diffuses with time, there is a positive correlation between the lateral spread (\sxy ) of the collected charge on the pixel array and the depth of the interaction ($z$).
After a user-defined exposure time, the charge collected in every pixel is transferred serially into a low-noise output node for measurement.
A CCD readout where columnwise segments of 100 pixels were grouped in a single charge measurement results in an image with 4116$\times$42 pixels.
The image contains a two-dimensional stacked history (projected on the $x$-$y$ plane) of all ionization produced throughout the exposure. 
For details on the readout of DAMIC CCDs see Ref.~\cite{Aguilar-Arevalo:2016ndq}.

The DAMIC detector was commissioned in the summer of 2017 with a red (780\,nm) light-emitting diode installed inside the cryostat.
Images were acquired with varying light exposures to confirm the efficient charge transfer and to calibrate the output signal of each CCD in units of eV electron equivalent (1\,$e^-$$=$3.8\,\eve) following the procedure in Ref.~\cite{Aguilar-Arevalo:2016ndq}.

For the dark matter search, data images were acquired with \powert{4}{3}\,s or \powero{5}\,s exposures, immediately followed by ``blank"  images whose exposure is solely the 130\,s readout time.
Image quality was monitored throughout data acquisition, including visual inspections.
Images with visible gradients from transients of leakage current after restarting the electronics or caused by temperature changes, or with visible patterns from readout noise, were discarded before processing.
Images acquired when there was a measurable level of radon ($>$5\,Bq\,m$^{-3}$) around the cryostat were also excluded because they have an increased background from penetrating $\gamma$ rays emitted by short-lived radon daughters.
A total of 5607 images from 801 exposures, together with their corresponding blanks, were considered for this analysis, with an integrated exposure time of 308.1\,d.

Image processing started with the pedestal removal and correlated-noise subtraction procedures of the DAMIC analysis pipeline described in Refs.~\cite{Aguilar-Arevalo:2016ndq} and~\cite{Aguilar-Arevalo:2016zop, *PhysRevLett.123.181802}, respectively.
We used the images for this analysis and images from higher-temperature data runs acquired in early 2017 to identify spatially localized regions of high leakage current due to lattice defects and generated ``masks'' following the procedure in Ref.~\cite{Aguilar-Arevalo:2016ndq}.
Additionally, pixels on the edges of the CCDs with coordinate $x$$\le$128 or $x$$>$3978, which exhibit transient leakage current following the restart of the electronics, were included in the masks.
The application of the masks removed 7\% of pixels and results in a distribution of pixel values centered at zero and dominated by white noise with \spix$\sim$1.6\,$e^-$.
Only 29 of the processed images have readout noise that is inconsistent with white noise, having at least one negative pixel with value $<$$-5$\,\spix , and were discarded.

\begin{figure}[t!]
	\centering
	\includegraphics[width=0.48\textwidth]{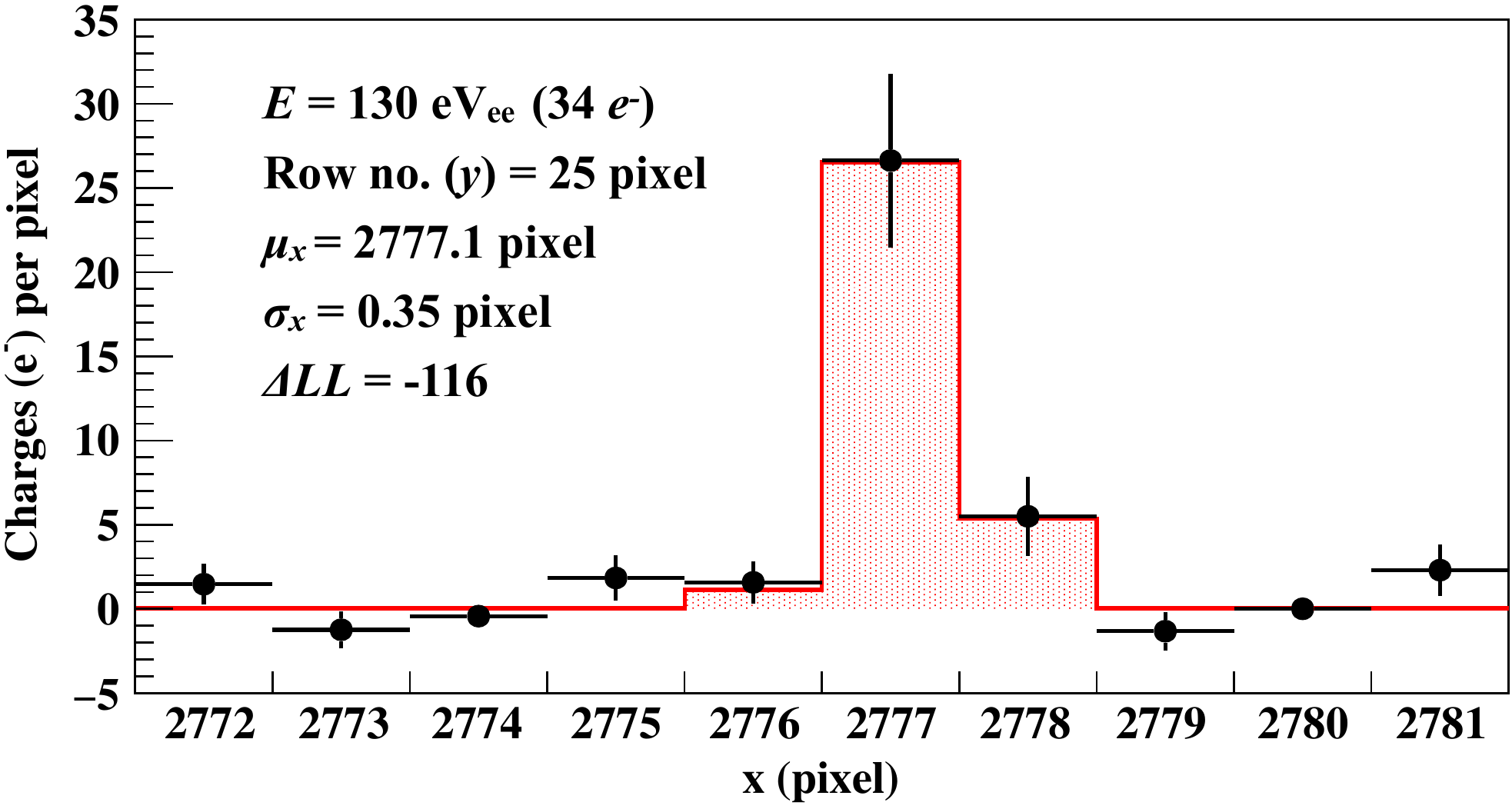}
	\caption{Data cluster in the WIMP search energy region. The black markers show the pixel values along the row while the red histogram is the result of the best-fit Gaussian function. Cluster variables are given in the inset.}
	\label{fig:event_eg}
\end{figure}

For the low-energy events of interest, the range of the ionizing particles is much smaller than the CCD pixel size and diffusion dominates the distribution of charge on the pixel array.
Because the charge was read out in columnwise segments 100 pixels high, information on the distribution of charge along the $y$ axis was lost.
Hence, the pattern on the image can be described by a Gaussian distribution along a row, whose amplitude is proportional to the deposited energy $E$, mean $\mu_x$ is the $x$ coordinate of the interaction, and \sx\ is the spatial width in the $x$ dimension.
We identified clusters of charge using both a ``fast'' algorithm, which groups contiguous pixels with signal larger than 4\,\spix , and a ``likelihood'' algorithm, which performs a statistical test in a moving window along a row to search for the preference of a Gaussian template over baseline white noise.
In addition, for the likelihood clusters we computed \dll , the result of a likelihood ratio test between the best-fit Gaussian function and a flat baseline, such that more negative values correspond to a higher statistical significance of the cluster.
Figure~\ref{fig:event_eg} shows an example of an identified low-energy cluster and the corresponding best-fit Gaussian function.
The clustering algorithms, and the accuracy and precision of event reconstruction are described in Ref.~\cite{Aguilar-Arevalo:2016ndq}.

The relation between \sx\ and the $z$ coordinate of an energy deposition in the CCD active region can be modeled as $\sigma_{x}^2$$=$$-A\ln |1-bz|$.
The values $A$$=$285$\pm$24\,\um$^2$ and $b$$=$\powert{-4}{(8.2$\pm$0.3)}\,\um$^{-1}$ were obtained from fits to straight cosmic-muon tracks acquired when the CCDs were characterized on the surface before deployment at SNOLAB.
The details on the diffusion model and the specifics of the calibration can be found in Ref.~\cite{Aguilar-Arevalo:2016ndq}.
A comparison between the observed maximum diffusion (\smax ) in our data set and its expected value from the diffusion relation showed a \%-level deviation proportional to $E$.
A correction was applied to the model to match the observed \smax\ in the data: \sx $=$$\sqrt{-A\ln |1-bz|}\times(\alpha+\beta E)$, with $\alpha$$=$0.956 and $\beta$$=$0.0059\,${\rm keV}_{\rm ee}^{-1}$.

To construct a radioactive background model, we performed a {\tt GEANT4}~\cite{ALLISON2016186} Monte Carlo simulation tracking the radioactive decay products of 23 isotopes in a detailed detector geometry consisting of 64 separate volumes~\cite{joao, *back_model}.
A custom simulation was used for the response of the CCDs, which includes models for charge generation and transport, pixelation and readout noise.
The fast clustering algorithm was run on the simulated events and data to obtain distributions in reconstructed $E$ and $\sigma_x$ for direct comparison.
The simulations were grouped to form 49 templates differing in event properties such as common decay chain or material origin.

We then performed a two-dimensional binned Poisson likelihood fit to the data from CCDs~2--7 with simulated $(E,\sigma_x)$ templates, reserving CCD~1 for a cross-check.
The fit was performed between 6 and 20\,k\eve , where the presence of a statistically significant WIMP signal has been excluded by previous silicon experiments~\cite{Agnese:2013rvf}.
We excluded clusters in which any pixel was touching a masked pixel, or whose shape was not well described by the best-fit Gaussian.
The energy region 7.5--8.5\,k\eve\ was not considered in the fit to exclude the $K$-shell line from copper fluorescence, a secondary atomic process that was outside the scope of this work to reproduce by means of {\tt GEANT4}.
The amplitude of each template was a parameter in the fit.
The activities of most isotopes were constrained by radioactive screening results, using Gaussian penalty terms in the likelihood function according to the uncertainty of each measurement.
The cosmogenic radioactivity of copper components was calculated from the history of the copper assuming surface activation rates from Ref.~\cite{laubenstein}.

\begin{figure}[t!]
	\centering
	\includegraphics[width=0.48\textwidth]{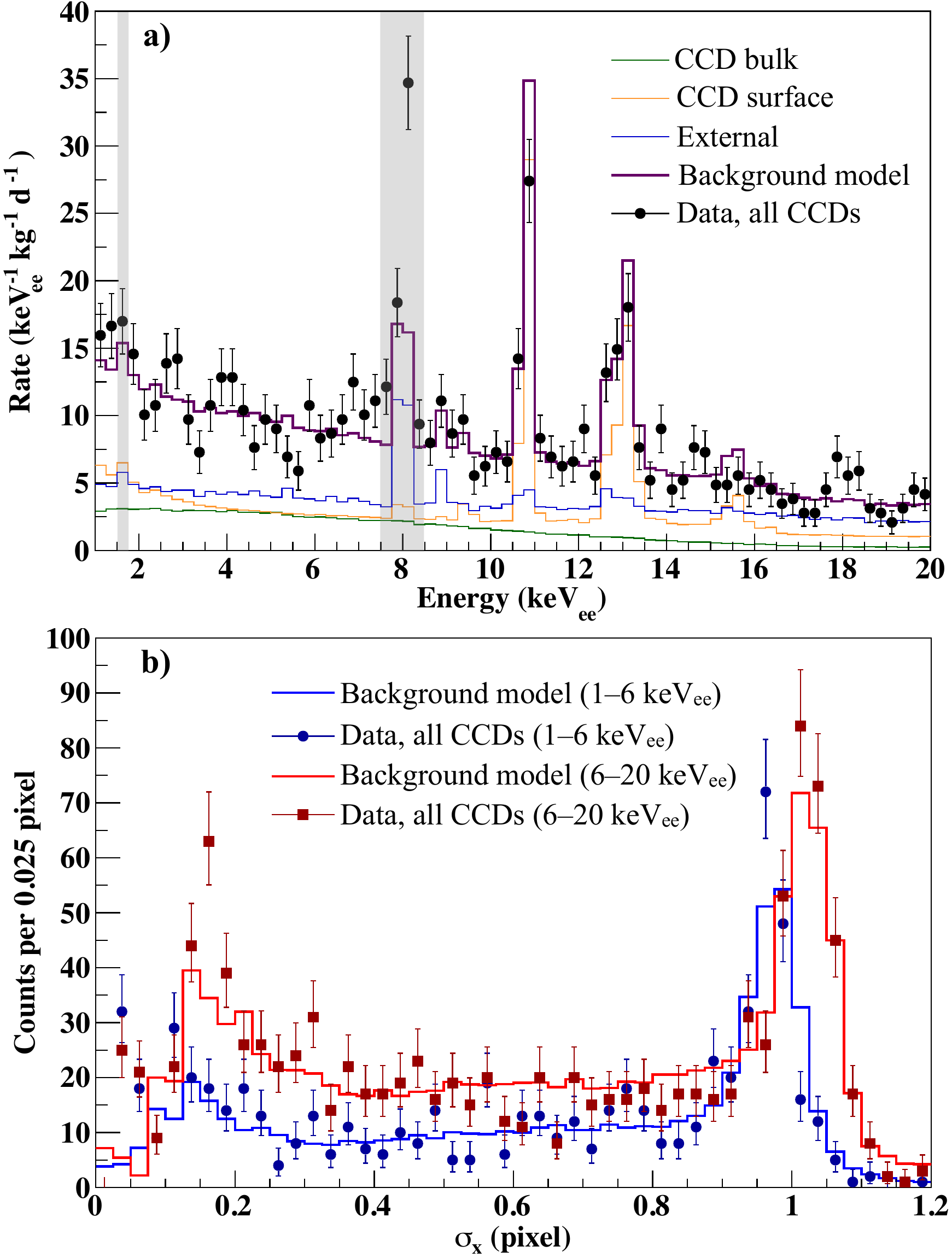}
	\caption{Projections in $E$ and \sx\ of the best-fit background model (solid lines) compared to the fast clustered data (markers). a)~The total background model spectrum is shown along with four separate contributions grouped by background origin. Shaded energy regions are excluded from the analysis. b)~Comparison of \sx\ distributions in the fit energy region (red) and at lower energies (blue). The peak at low (high) \sx\ corresponds to events at the front (back) of the CCDs.
	}
	\label{fig:background_model}
\end{figure}

We present the fit results for all CCDs combined and projected on the $E$ and \sx\ dimensions in Fig.~\ref{fig:background_model}, together with the extrapolation of the best-fit background model in the 1--6\,k\eve\ range.
The background model is statistically consistent with the spectra observed by CCDs 2--7 individually, and correctly predicts the 50$\%$ lower background measured by CCD~1.
A dominant component (3.8$\pm$0.4\,\dru\ in the range 1--6 k\eve ) is from the decay of \pbten\ (and daughter \biten ) on the surfaces of the CCDs.
This contamination comes from radon exposure during storage and handling of the devices, including contamination on the surface of the wafer before fabrication, now buried 3\,\um\ from the surfaces of the CCDs.
The bulk component (2.9$\pm$0.7\,\dru ) mostly comes from the decays of \tritium\ and $^{22}$Na from the cosmogenic activation of the silicon while the CCDs were on the surface, with a subdominant contribution (0.17$\pm$0.03\,\dru ) from \sitwo\ (and daughter \ptwo ), constrained from the number of \sitwo-\ptwo\ spatial coincidences observed in the data following the strategy from Ref.~\cite{Aguilar-Arevalo:2015lvd}.
The background from energetic electrons and photons external to the CCD (4.4$\pm$0.5\,\dru ) comes from cosmogenic cobalt and \pbten\ contamination in the copper components of the detector, as well as uranium and thorium in the Kapton flex cables that connect to the CCDs.

The main systematic uncertainty in our analysis is related to the presence of an $\sim$5\,\um-thick partial charge collection (PCC) region in the back of the CCDs ($z$$\sim$670\,\um ) caused by diffusion of phosphorous (P) from the highly doped back-side electrical contact into the lightly doped CCD active region.
At intermediate P concentrations, some of the charge generated by ionization events recombines before diffusing into the active region, leading to PCC events.
To model this transition, we simulated at different depths charge packets under diffusion and accounted for charge losses from recombination using the charge mobility and lifetime measurements as a function of P concentration from Ref.~\cite{DELALAMO19871127, *tdepmob}.
The P concentration was obtained by secondary-ion mass spectrometry of the CCD back side.
We considered a discrete set of variations within their uncertainties from the nominal model and in each case simulated the response of the CCD to back surface \pbten\ (and \biten ) decays.
The simulated spectra are almost identical in the 6--20\,k\eve\ energy range and cannot be distinguished by the background model fit but lead to significantly different spectra at low energies.
We found that differences between the simulated spectra for different PCC-model variations and specific locations of the \pbten\ contamination can be parametrized by the functional form $C\exp(-\sqrt{[E/\mbox{keV$_{\rm ee}]$}}/0.18)$, with $C$ being dominantly dependent on the relative depth of the \pbten\ source and the point at which the charge collection probability becomes $>$0.
Thus, we consider this functional form as a correction to our background model to account for the systematic uncertainty in the details of the PCC region.

The likelihood clustering output was used to search for any event excess in the energy range 0.05--6\,k\eve\ over the prediction by the background model.
Images with average pixel charge $>$0.47\,$e^-$ were excluded due to their high levels of shot noise associated with transients of leakage current following the restart of the electronics or LED illumination for CCD calibration. This results in a final target exposure of 10.93\,kg$\,$d.
We selected clusters that were not touching the mask or another cluster, and whose pixel-value distributions were well described by the Gaussian fit.
A selection on \dll\ as in Ref.~\cite{Aguilar-Arevalo:2016ndq} was then used to reject clusters compatible with noise.
We started with blank images, which contain only readout noise, and added leakage charge according to the value measured in the corresponding exposed image.
The likelihood clustering algorithm was run on the simulated images to obtain a sample of simulated clusters.
We determined from the \dll\ distribution of all simulated images that a selection of \dll $\leq$$-22$ results in $<$0.1 clusters from noise in the final dataset.

The detection efficiency for ionization events as a function of energy was estimated with the CCD-response Monte Carlo following the procedure that was validated with $\gamma$-ray calibration data in Ref.~\cite{Aguilar-Arevalo:2016ndq}.
We simulated pointlike ionization events with uniform distributions in energy and depth ($z$) and added the pixel values directly on the blank images.
The likelihood clustering algorithm was run, and from the fraction of simulated clusters of a given energy that survive the selection criteria, we reconstructed the acceptance for ionization events as a function of energy.
The acceptance starts at 10$\%$ at 50\,\eve , increasing to 50$\%$ at 77\,\eve , until it plateaus at 90$\%$ above 120\,\eve\ because of the fraction of clusters that touch the mask.

\begin{figure}[t!]
	\centering
	\includegraphics[width=0.48\textwidth]{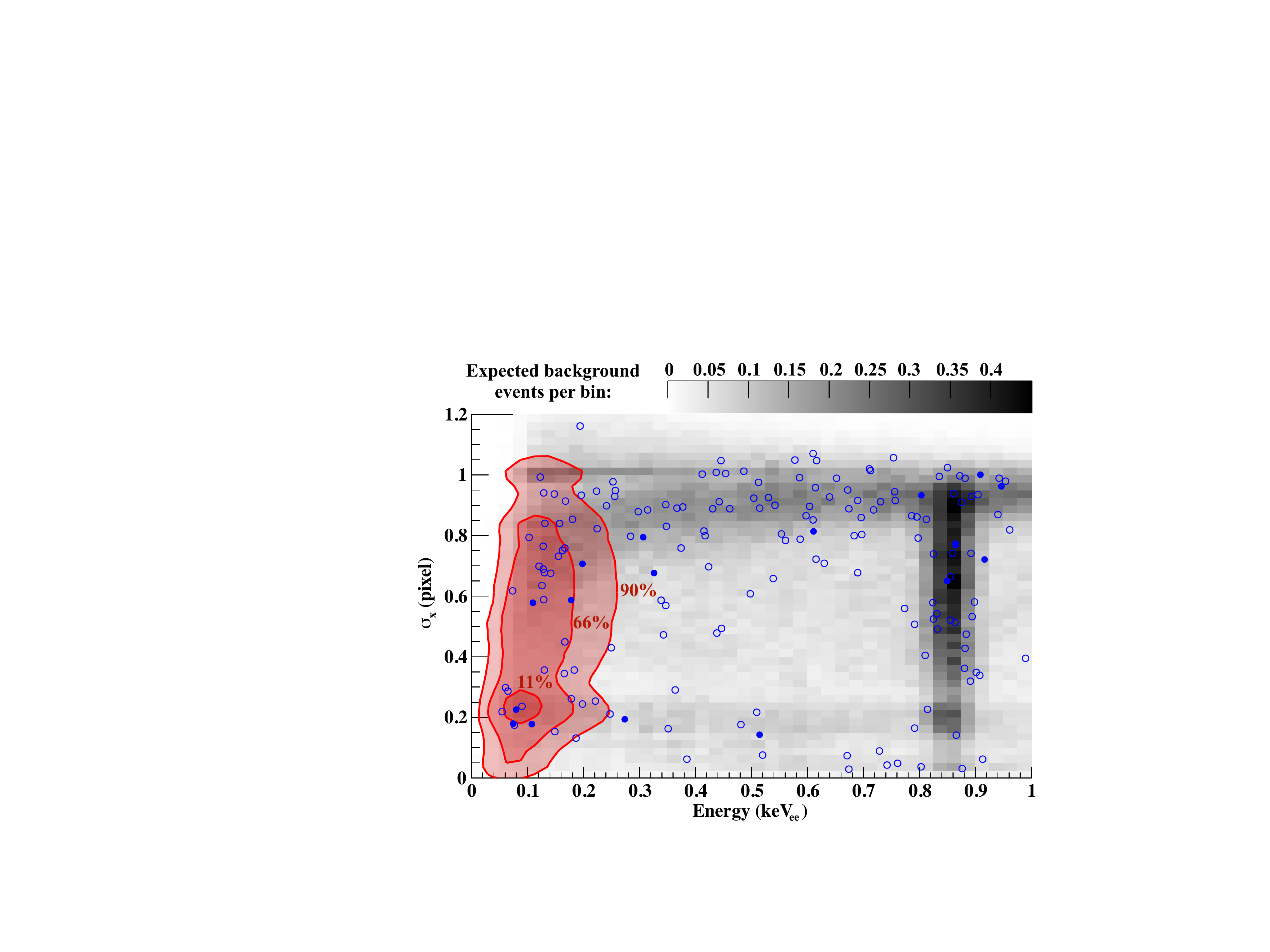}
	\caption{Best fit to the final likelihood clustered data sample in $(E,\sigma_x)$ space. The blue data points are overlaid on the best-fit background model in gray. Open circles correspond to CCDs 2--7, while filled circles correspond to CCD 1. The red contours represent the best-fit exponential excess with $\epsilon$$=$67\,\eve .
	The Ne $K$ deexcitation line (0.87\,k\eve ) emitted following electron capture by $^{22}$Na in the CCD bulk is visible.
	}
	\label{fig:final_fit}
\end{figure}

To obtain background predictions that can be compared to the likelihood clustering output, we produced images with simulated events sampled randomly from the $(E,z)$ templates of the baseline background model and an additive systematic correction to account for the PCC region on the back side, treating CCD~1 and CCDs~2--7 separately.
We applied the same clustering, reconstruction, and selection procedure as in the data to construct probability density functions (PDFs) in $(E,\sigma_x)$ space normalized to 1 in the fit region $E\in[0.05, 6] \ \rm keV_{ee}$ and $\sx \in [0, 1.2]$ pixel, excluding Si $K$ fluorescence $E\in[1.6, 1.8] \ \rm keV_{ee}$.
For a statistical test to check the consistency between the background model and the data, we assumed a decaying exponential with characteristic decay constant $\epsilon$ convolved with the detector energy response as a generic signal PDF obtained from the $(E,\sigma_x)$ template of uniformly distributed events in $(E,z)$ space by scaling the amplitude as a function of energy and normalizing to 1 in the fit region.
We defined a two-dimensional $(E, \sx)$ unbinned extended likelihood function following the formalism in Ref.~\cite{Aguilar-Arevalo:2016ndq}.
Clusters from CCD~1 and CCDs~2--7 were considered independent datasets with their own background PDFs.
We minimized the joint $-\ln \mathcal{L}$ using {\tt MINUIT} with the PDF amplitudes $b_{1,2\text{--}7}$, $c$, and $s$ (baseline background, PCC correction and generic signal), and $\epsilon$ as free parameters.
We included Gaussian constraints on $b_{1,2\text{--}7}$ according to the uncertainty in the amplitude of the background model above 6\,k\eve .
Our best fit exhibits a preference for an exponential bulk component with $s$$=$17.1$\pm$7.6 events and decay constant $\epsilon$$=$67$\pm$37\,\eve .
The best-fit value for $c$ corresponds to a distance between \pbten\ contamination on the back side of the original wafer and the start of charge collection of $0.75^{+0.50}_{-0.35}$\,\um , consistent with results from \ironfive\ x-ray calibrations~\cite{PCC}.
Figure~\ref{fig:final_fit} shows the data clusters overlaid on the background model, with contours delimiting the preferred bulk component.
We estimated a goodness-of-fit $p$ value of 0.10 by running our fit procedure on Monte Carlo samples drawn from the best-fit PDF.
A likelihood ratio test between the best-fit result and the background-only hypothesis ($s$$=$0) disfavors the background-only hypothesis with a $p$ value of $2.2\times10^{-4}$.
If we perform the fit to the data from CCD~1 or CCDs~2--7 separately, the resulting bulk component is statistically consistent between the two datasets with a higher statistical significance in CCD~1, which has the lowest background.

The statistical significance of the exponential bulk component is driven by an excess of events at low energies with \sx$\sim$0.2\,pixel.
We explored the possibility that these events arise from an improper modeling of front-surface ($z$$\sim$0) events, which can also populate this region of parameter space.
We removed from the data and in the generation of the PDFs clusters where only one pixel has a value greater than 1.6\,\spix , which correspond to 56$\%$ of front-surface events but only 6.5$\%$ of bulk events with energies $<$200\,\eve .
A fit performed to the data following this selection returns values for $s$ and $\epsilon$ consistent with the previous result, with an increased $p$ value of $2.6\times10^{-3}$.

\begin{figure}[t!]
	\centering
	\includegraphics[width=0.48\textwidth]{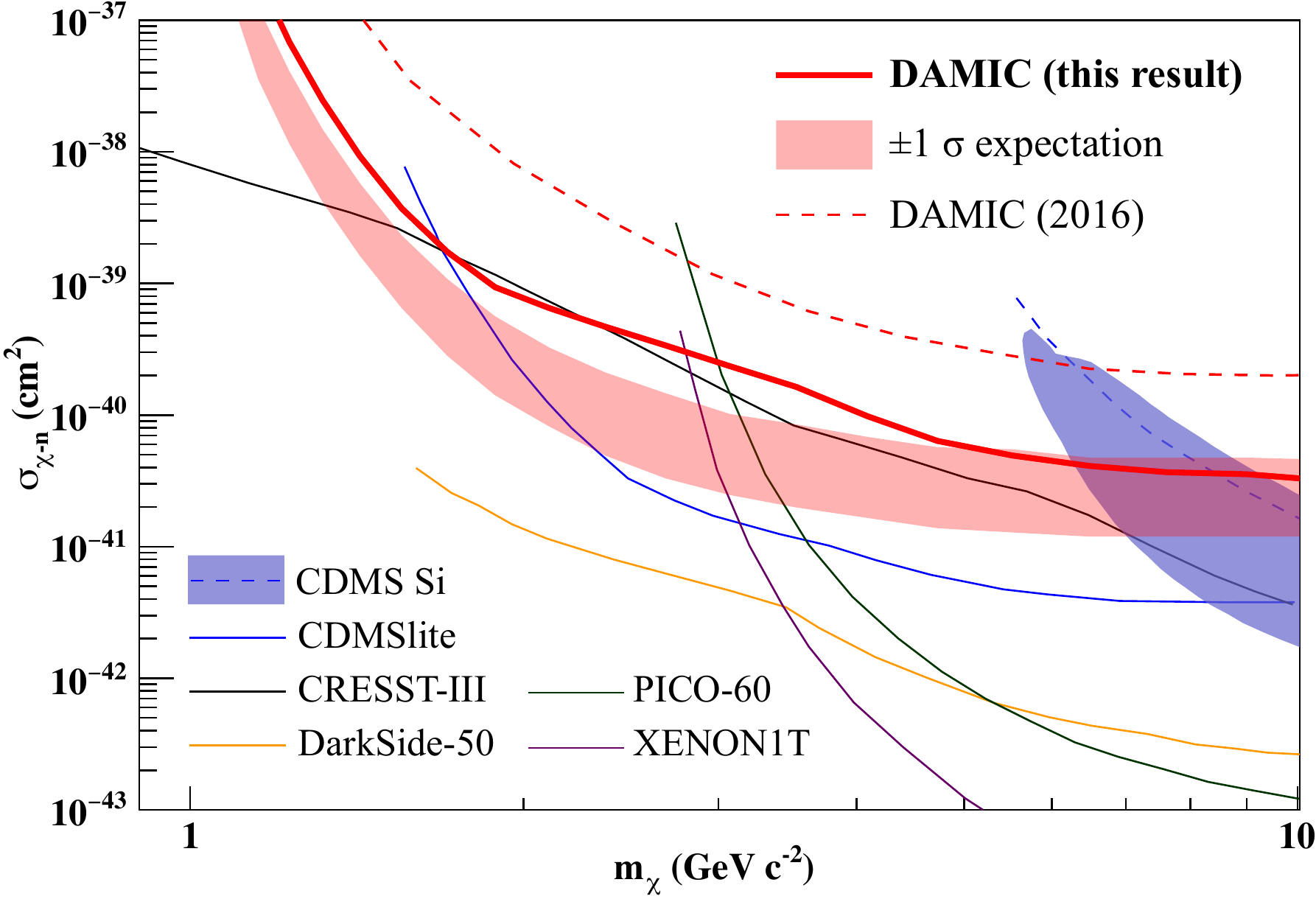}
	\caption{Upper limit (90\% C.L.) on \swn\ obtained from this analysis (solid red line). The expectation $\pm$$1\,\sigma$ band if only known backgrounds are present in our data set is shown by the red band. For comparison, we also include 90\% C.L. exclusion limits from our previous result with a 0.6\,kg\,d exposure~\cite{Aguilar-Arevalo:2016ndq}, other experiments~\cite{Agnese:2013rvf, darkside50, *cdmslite, *pico60, *cresstiii, *xenon1ts2}, and the 90\% C.L. contours for the WIMP-signal interpretation of the CDMS silicon result~\cite{Agnese:2013rvf}.}
	\label{fig:exclusion_limit}
\end{figure}

Limited statistics and possible unidentified inaccuracies in the detector background model prevent a definite interpretation of this event excess.
We plan to further investigate its origin by improving the measurement of the ionization spectrum with lower noise skipper CCDs~\cite{skipper} deployed in the DAMIC cryostat at SNOLAB.
Nevertheless, we set upper limits on the amplitude of a signal from spin-independent coherent WIMP-nucleus elastic scattering.
Starting from the $(E,\sigma_x)$ template of uniformly distributed events, we constructed a PDF of a WIMP signal by scaling the amplitude as a function of energy by the expected spectrum from nuclear recoils~\cite{LEWIN199687} for a given \mchi . We used a speed distribution with standard galactic halo parameters: escape speed of 544\,km\,s$^{-1}$, most probable Galactic WIMP speed of 220\,km\,s$^{-1}$, mean orbital speed of Earth with respect to the Galactic Center of 232\,km\,s$^{-1}$, and local WIMP density of 0.3\,\gev\,cm$^{-3}$.
To translate from nuclear-recoil energy to the measured electron-equivalent energy, we used the parametrization from Ref.~\cite{Aguilar-Arevalo:2016ndq} based on neutron calibrations~\cite{Chavarria:2016xsi, *Izraelevitch_2017} that cover the nuclear-recoil energy range 0.7--20\,keV (0.06--7\,k\eve ) with a linear extrapolation toward lower energies that results in no ionization below 0.3\,keV.
We included in our fit function a WIMP-signal PDF with \mchi , and performed the fit with \swn\ free.
From likelihood ratio tests between this best-fit result and the result of a constrained fit with fixed \swn , we calculated the statistical significance for the WIMP signal in $(m_\chi, \sigma_{\chi-n})$ space.
Figure~\ref{fig:exclusion_limit} shows the 90$\%$ C.L. upper limit obtained from our data compared to other experiments.
We also present the $\pm$1\,$\sigma$ expectation band by running the limit-setting procedure on Monte Carlo datasets drawn from our best-fit background model, in the absence of the unknown bulk component.

The derived exclusion limit is the most stringent from a silicon target experiment for WIMPs with \mchi $<$9\,\gev .
Although the presence of the unknown bulk component causes a mismatch between the derived and expected upper limit at small \mchi , the agreement for \mchi $>$6\,\gev\ implies that the observed excess is inconsistent with the standard WIMP-signal interpretation of the nuclear-recoil event excess from the CDMS silicon experiment~\cite{Agnese:2013rvf}.
Consequently, we excluded with the same nuclear target a significant fraction of the parameter space that corresponds to this interpretation.
Generally, this result uncovers with a sizeable exposure the ionization spectrum in silicon down to nuclear-recoil energies of 0.6\,keV, an order-of-magnitude improvement from the 7\,keV threshold of the CDMS experiment, providing a direct constraint for any dark matter interpretation of the CDMS excess.

\begin{acknowledgments}
We are grateful to SNOLAB and its staff for support through underground space, logistical, and technical services. SNOLAB operations are supported by the Canada Foundation for Innovation and the Province of Ontario Ministry of Research and Innovation, with underground access provided by Vale at the Creighton mine site.
The CCD development work was supported in part by the Director, Office of Science, of the U.S. Department of Energy under Contract No. DE-AC02-05CH11231.
We acknowledge financial support from the following agencies and organizations:
National Science Foundation through Grant No. NSF PHY-1806974 and Kavli Institute for Cosmological Physics at The University of Chicago through an endowment from the Kavli Foundation;
Gordon and Betty Moore Foundation through Grant No. GBMF6210 to the University of Washington;
Fermi National Accelerator Laboratory (Contract No. DE-AC02-07CH11359);
Institut Lagrange de Paris Laboratoire d'Excellence (under Reference No. ANR-10-LABX-63) supported by French state funds managed by the Agence Nationale de la Recherche within the Investissements d'Avenir program under Reference No. ANR-11-IDEX-0004-02;
Swiss National Science Foundation through Grant No. 200021\_153654 and via the Swiss Canton of Zurich;
Mexico's Consejo Nacional de Ciencia y Tecnolog\'{i}a (Grant No. 240666) and  Direcci\'{o}n General de Asuntos del Personal Acad\'{e}mico--Universidad Nacional Aut\'{o}noma de M\'{e}xico (Programa de Apoyo a Proyectos de Investigaci\'{o}n e Innovaci\'{o}n Tecnol\'{o}gica Grants No. IB100413 and No. IN112213);
STFC Global Challenges Research Fund (Foundation Awards Grant No. ST/R002908/1).
\end{acknowledgments}

\bibliography{myrefs.bib}

\end{document}